\documentclass{article}
\usepackage{arxiv}
\pdfoutput=1

\usepackage[utf8]{inputenc} % allow utf-8 input
\usepackage[T1]{fontenc}    % use 8-bit T1 fonts
\usepackage{hyperref}       % hyperlinks
\usepackage{url}            % simple URL typesetting
\usepackage{booktabs}       % professional-quality tables
\usepackage{amsfonts}       % blackboard math symbols
\usepackage{nicefrac}       % compact symbols for 1/2, etc.
\usepackage{microtype}      % microtypography
\usepackage{lipsum}		% Can be removed after putting your text content
\usepackage{graphicx}
\usepackage{natbib}
\usepackage{doi}
\usepackage{amsmath}
\usepackage{threeparttable}
\usepackage{multirow}

\newcommand{\figref}[1]{Fig.~\ref{#1}}                      
\newcommand{\tabref}[1]{Table~\ref{#1}}                      
\newcommand{\eqnref}[1]{Eq. \ref{#1}}

\title{Streaming Lossless Volumetric Compression of Medical Images Using Gated Recurrent Convolutional Neural Network}

\author{ \href{https://orcid.org/0000-0003-2782-9217}{\includegraphics[scale=0.06]{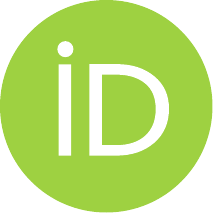}\hspace{1mm}Qianhao Chen} \\
	College of Biomedical Engineering \\
	Zhejiang University\\
	\texttt{chen\_qh@zju.edu.cn} \\
	\And
	{Jietao Chen} \\
	College of Biomedical Engineering \\
	Zhejiang University\\
	\texttt{chenjietao1514@163.com} \\
}

\renewcommand{\shorttitle}{\textit{arXiv}}

\hypersetup{
pdftitle={Streaming Lossless Volumetric Compression of Medical Images Using Gated Recurrent Convolutional Neural Network},
pdfsubject={q-bio.NC, q-bio.QM},
pdfauthor={Qianhao Chen, Jietao Chen},
pdfkeywords={Volumetric image compression, real-time system, gate mechanism},
}

\begin{document}
\maketitle

\begin{abstract}
Deep learning-based lossless compression methods offer substantial advantages in compressing medical volumetric images.
Nevertheless, many learning-based algorithms encounter a trade-off between practicality and compression performance.
This paper introduces a hardware-friendly streaming lossless volumetric compression framework, utilizing merely one-thousandth of the model weights compared to other learning-based compression frameworks.
We propose a gated recurrent convolutional neural network that combines diverse convolutional structures and fusion gate mechanisms to capture the inter-slice dependencies in volumetric images.
Based on such contextual information, we can predict the pixel-by-pixel distribution for entropy coding.
Guided by hardware/software co-design principles, we implement the proposed framework on Field Programmable Gate Array to achieve enhanced real-time performance.
Extensive experimental results indicate that our method outperforms traditional lossless volumetric compressors and state-of-the-art learning-based lossless compression methods across various medical image benchmarks.
Additionally, our method exhibits robust generalization ability and competitive compression speed.
\end{abstract}

% keywords can be removed

% =============================================================
\section{Introduction}
\label{sec:introduction}
The advancement of medical imaging technologies has resulted in the generation of massive medical volumetric images, including Computed Tomography (CT) and Magnetic Resonance Imaging (MRI).
These volumetric images play a crucial role in clinical diagnosis owing to their detailed anatomical structures \citep{fields2021imaging, lenchik2019automated, tang2020whole}.
However, the high resolution and expanded scanning range, providing richer information, pose substantial challenges to hardware storage and transmission.
In contrast to natural images, distortion in medical images resulting from lossy compression can affect the precision of analytical tasks \citep{liu2017current}.
Consequently, there is an urgent need to explore effective lossless compression techniques for 3-D medical images.
In recent years, numerous methods have been proposed for the lossless compression of volumetric images, which can be roughly divided into three categories: transform-based, video-based, and deep learning-based.

Medical imaging systems typically adhere to the guidelines established by the Digital Imaging and Communications in Medicine (DICOM) organization \citep{mustra2008overview}.
These guidelines specify the utilization of major image coding standards, including PNG \citep{roelofs1999png}, Lossless JPEG (JPEG-LS) \citep{weinberger2000loco}, JPEG-2000 \citep{taubman2002jpeg2000}, etc. 
These methods employ either Discrete Cosine Transformation (DCT) or Discrete Wavelet Transformation (DWT) to convert the input image from the pixel domain to the more energy-concentrated frequency domain.
Furthermore, the 3-D extension of JPEG-2000, known as JP3D \citep{bruylants2009jp3d}, is widely utilized for losslessly compressing volumetric images.
However, transform-based methods rely on hand-crafted transform basis and cannot be optimized according to specific data characteristics, thereby limiting their performance.

Other researchers have explored video-based methods.
They treat medical volumetric images as time-continuous slices, estimating a motion field to leverage temporal correlation in the time dimension.
High Efficiency Video Coding (HEVC) \citep{sullivan2012overview}, commonly referred to as H.265, utilizes intra-prediction and motion compensation to enhance compression performance.
FF Video 1 (FFV1) \citep{niedermayer2013ffv1} enables the context model to adapt over multiple frames.
Due to the substantial size of the context table, it proves beneficial for compression.
However, the inter-slice correlation pattern in medical volumetric images differs from the inter-frame correlation pattern in typical videos \citep{kuanar2019low}.
Consequently, it cannot effectively utilize inter-frame prediction to exploit inter-slice redundancy in medical volumetric images.

Recently, owing to the substantial advancement of Deep Neural Networks (DNNs), numerous deep learning-based methods have been introduced and have attained state-of-the-art compression ratios.
Learning-based methods, such as PixelCNN++ \citep{salimans2017pixelcnn++}, L3C \citep{mentzer2019practical}, VQ-VAE \citep{kang2022pilc}, IDF \citep{hoogeboom2019integer}, etc., utilize generative models for image prediction, and the compression ratio relies on how accurate the prediction is \citep{nguyen2021lossless}.
Furthermore, to exploit the inter-slice redundancy of medical volumetric images, Chen \textit{et al.} introduced an intra-slice and inter-slice conditional entropy coding model (ICEC) by adding a latent feature buffer of previous image slices \citep{chen2022exploiting}.
Additionally, notable methods include aiWave \citep{xue2022aiwave}, a method based on the 3-D trained affine wavelet-like transform, as well as extensions of 3-D autoencoder \citep{gao2020volumetric} and 3-D Minimum Rate Predictors (MRP) \citep{lucas2017lossless} methods, which directly address volumetric data to eliminate inter-slice redundancy.

While deep learning-based methods optimize end-to-end using massive training data, they still have the following limitations:
\begin{itemize}
\item{Generative models have been shown to perform poorly in high-bitrate scenarios \citep{helminger2020lossy}.
Regrettably, medical images typically have a high bit depth, exacerbating this issue.}
\item{Current methods struggle to capture dependencies between slices of medical volumetric images effectively, impeding the reduction of inter-slice redundancy.}
\item{The majority of learning-based methods utilize large models with significant computational complexity, thereby restricting their hardware portability and compression speed.
Considering a trade-off between practicality and model complexity is essential \citep{nakkiran2021deep}.}
\end{itemize}

In deep learning, it has been proven that large models have the most significant improvement on examples where the small model is most uncertain \citep{narayan2022predicting}.
Nevertheless, in compression tasks, a few pixels with uncertain predictions exert minimal influence on the compression ratio.
Consequently, we introduced a lightweight Streaming Recurrent Lossless Volumetric Image Compression System (SR-LVC) designed explicitly for medical volumetric images.
Specifically, we proposed a gated recurrent convolutional neural network that combines Masked Convolution Neural Networks (Masked-CNN) 
\citep{van2016conditional} and fusion gate mechanisms to eliminate intra-slice and inter-slice redundancy.
Leveraging the latent features of the current pixel and the hidden states containing inter-slice information, the fusion gate utilizes multiple gating mechanisms to merge hidden states and latent features.
Lastly, the probability estimator provides the distribution of the current encoding based on the fusion state.
Overall, our lossless compression framework optimally exploits both intra- and inter-slice redundancy with a remarkably low number of model weights----specifically, a mere 4866.

Our contributions can be summarized as follows:
\begin{itemize}
\item{The proposed fusion gate enables selective retention or forgetting of the information, making it more effective for capturing inter-slice dependencies. 
In contrast to the explicit transformation-based methods (\textit{e.g.}, 3-D wavelet \citep{xue2022aiwave}) and other gating mechanisms (\textit{e.g.}, ICEC \citep{chen2022exploiting}), it exhibits lower computational complexity and is better suited for streaming implementation.}
\item{The proposed compression framework is a Fully Convolutional Network (FCN).
Compared to other learning-based compression methods, our approach supports full resolution and high bit-depth input, expanding the range of application scenarios.}
\item{To the best of our knowledge, it is the first streaming learning-based framework for lossless volumetric image compression.  
The fewer model weights (only $4866$ parameters) make the proposed method more hardware-friendly, facilitating the streaming implementation on Field Programmable Gate Array (FPGA).}
\item{We conduct extensive experiments on various volumetric image datasets.
The results indicate that our method's compression performance surpasses that of state-of-the-art lossless volumetric image compression methods.
Furthermore, our method demonstrates robust generalization ability and competitive compression speed.}
\end{itemize}

This paper is organized as follows.
Section II reviews deep learning-based methods for lossless volumetric image compression.
Section III provides a detailed description of our streaming framework (SR-LVC).
Section IV illustrates the hardware architecture and provides details on FPGA implementation.
Section V presents the experimental results and analysis.
Section VI concludes the entire paper.
% =============================================================

% =============================================================
\section{Related Work}
\label{sec:headings}
Data compression has an impact on data accessibility.
The total Data Access Time (DAT) is determined by compression time, writing time, and startup latency \citep{zhang2017realizing}.
Categorized by startup latency, the learning-based volumetric compressors can be roughly divided into three groups: volume-wise methods, slice-wised methods, and pixel-wised methods.

\subsection{Volume-wise Methods}
Volume-wise methods entail the preparation of the entire volumetric image before compression begins, resulting in an extended startup time interval.
Typically, these methods employ 3-D explicit transformations (\textit{e.g.}, 3-D convolutions, 3-D wavelet) to eliminate spatial redundancy within the local receptive field.

For instance, Gao \textit{et al.} utilized a 3-D convolution autoencoder model to compress 3-D image cubes and integrated non-local attention models to jointly capture local and global correlations \citep{gao2020volumetric}.
Additionally, Xue \textit{et al.} devised a 3-D trained wavelet-like transform for an end-to-end compression framework named aiWave \cite{xue2022aiwave}. 
This framework primarily comprises an affine wavelet-like transform module and a context-based entropy coding module utilizing 3-D convolution methods.
While experimental results demonstrated that aiWave achieved an optimal compression ratio, it still exhibited some limitations.
Specifically, owing to the complexity of the affine wavelet-like transform module, aiWave had 695.5MB weights and required an average of 900 seconds to process one slice.

\subsection{Slice-wise Methods}
Slice-wise methods commonly employ autoencoders to extract multi-scale features from individual slices.
Mentzer \textit{et al.} introduced a lossless image compression system named L3C, employing autoencoders to extract hierarchical feature representations \citep{mentzer2019practical}.
Additionally, Kang \textit{et al.} utilized a lightweight VQ-VAE \citep{van2017neural} to extract discrete latent feature representations \citep{kang2022pilc}.

The previously mentioned methods fail to leverage inter-slice correlation to eliminate spatial redundancy.
In response to this challenge, Chen \textit{et al.} introduced ICEC, utilizing gate mechanisms to retain information from previous slices \citep{chen2022exploiting}.
However, ICEC extracts multi-scale latent features before fusing them with gate mechanisms. 
It struggles to capture differences between slices effectively.
Moreover, due to the use of gate mechanisms with weights, ICEC deviates from being a Fully Convolutional Network (FCN) like L3C, making it unable to accept inputs of arbitrary resolutions.

\subsection{Pixel-wise Methods}
Pixel-wise methods enable compression to commence immediately upon the pixel's arrival, minimizing the startup time.
Inspired by the linear prediction filter, Google DeepMind first proposed Masked-CNN in PixelCNN \citep{salimans2017pixelcnn++}, modeling the discrete probability pixel by pixel.
Likewise, to eliminate spatial redundancy, Lucas \textit{et al.} introduced 3-D MRP \citep{lucas2017lossless}, utilizing a pyramid pixel sampling strategy from previous slices to estimate the current pixel.

Nevertheless, because of the limited expressive power of single-layer networks \citep{lu2017expressive}, pixel-wise methods frequently stack multiple layers to enhance the compression ratio.
Consequently, even though pixel-wise methods are theoretically capable of streaming implementation, their model complexity hinders efficient deployment on hardware, leading to a slower compression speed.

The SR-LVC proposed in this paper is classified as a pixel-wise method, and its lightweight design renders it hardware-friendly.
Additionally, it integrates a gated recurrent convolutional neural network to capture inter-slice dependencies, leading to improved compression performance and practicality compared to other learning-based methods.

% =============================================================
% Table
\renewcommand{\arraystretch}{1.5}
\begin{table}
\caption{Comparison of learning-based methods}
\label{tab:compare}
    \begin{center}
        \begin{tabular}{c c c c c }
        \hline
           & \textbf{aiWave} & \textbf{L3C} & \textbf{ICEC} & \textbf{SR-LVC} \\
        \hline
        Full Resolution & $\times$ & $\checkmark$ & $\times$ & $\checkmark$ \\
        High Bitrate & $\checkmark$ & $\times$ & $\checkmark$ & $\checkmark$ \\
        Inter Slice &  $\checkmark$ & $\times$ & $\checkmark$ & $\checkmark$ \\
        Startup Latency & Volume & Slice & Slice & Pixel \\
        Parameters & 699.5MB & 5M & 17M & 4.8K \\
        Hardware Implementation  & $\times$ & $\times$ & $\times$ & $\checkmark$ \\
        
        \hline
        \end{tabular}
    \end{center}
\end{table}
% =============================================================

\tabref{tab:compare} compares SR-LVC with other learning-based methods.
The columns represent whether the full-resolution image input is supported, whether high-bit depth is supported, the method's capability to eliminate inter-slice redundancy, the method's startup latency, the number of model weights, and whether it can be implemented in hardware.

% =============================================================

% =============================================================
\section{Proposed Framework}
This section will initially describe the principles of lossless volume compression.
Subsequently, we will present an overview and specific implementation details of our streaming framework.

\subsection{Principles}
Let $\mathbf{V}=(S_1,S_2,\ldots,S_T)$ represent the volumetric data (\textit{e.g.}, the MRI or CT data) with $T$ slices, while $S_t$ represents the $t^{th}$ slice of the volumetric data, with each slice consisting of ${H}\times{W}$ sub-pixels.
Inspired by the linear predictor, Masked-CNN uses previous neighboring pixels to predict the probability distribution of the current pixel:
\begin{equation}
\label{eq:1}
    p(x_{i,j}) = p(x_{i,j} \mid x_{i-K,j-K}, \ldots, x_{i,j-1}), x \in S_t.
\end{equation}
Therefore, the joint distribution of pixels over a slice $S_t$ is as the following product of conditional distribution:
\begin{equation}
    P(S_t) = \prod_{i,j}^{H,W}{p(x_{i,j})} = \prod_{i,j}^{H,W}{p(x_{i,j} \mid x_{i-K,j-K}, \ldots, x_{i,j-1})}.
\end{equation}
According to the Information Theory \citep{shannon1948mathematical}, the lower bound of the bitrate when losslessly compressing $S_t$ is the Shannon entropy:
\begin{equation}
    E(S_t) = \mathbb{E}_{S_t \sim P}[-log_2{P(S_t)}] = \mathbb{E}_{S_t \sim P}[-\sum_{i,j}^{H,W}log_2{p(x_{i,j}})].  
\end{equation}
This implies that the compression ratio is contingent upon the accuracy of the prediction.

Moreover, we can utilize the information from previous slices to aid in predicting the current slice.
We assume that the hidden state $\mathbf{h}_{t-1}$ contains information from the previous slices:
\begin{equation}
    \mathbf{h}_{t-1} = \mathcal{G}(\mathbf{h}_{t-2},S_{t-1}).
\end{equation}
The latent features $\mathbf{x}$ represent the previous pixels' information within the current slice:
\begin{equation}
    \mathbf{x}_{i,j} = \mathcal{F}(x_{i-K,j-K}, \ldots, x_{i,j-1}), x \in S_t.
\end{equation}
In this way, \eqnref{eq:1} becomes:
\begin{equation}
    p(x_{i,j}) = p(x_{i,j} \mid \mathbf{x}_{i,j}, \mathbf{h}_{t-1}).
\end{equation}
The probability distribution of the current pixel is influenced by both intra-slice and inter-slice information.
Formulating an appropriate function $\mathcal{G}$ facilitates capturing dependencies between slices, thereby enhancing prediction.

However, estimating the exact probability $p(x_{i,j})$ in practice is challenging.
Drawing inspiration from existing works \citep{mentzer2019practical, salimans2017pixelcnn++}, our goal is to estimate another probability function $\hat{p}$ by minimizing cross-entropy for data compression.
It is commonly assumed that the $\hat{p}$ follows a logistic distribution.
The Cumulative Distribution Function (CDF) is expressed as follows:
\begin{equation}
    CDF_{\hat{p}}(x, \mu, s) = \frac{1}{1 + e^{-(x-\mu)/s}},
\end{equation}
where $\mu$ is the location parameter, representing the distribution's mean, and $s$ is the scale parameter, representing the distribution's standard deviation. 
Varying values of $(\mu, s)$ can govern the location and shape of the logistic distribution.
Consequently, the probability estimator $\mathcal{P}$ only needs to output the $(\mu, s)$ value of the current pixel.
In image compression tasks, the minimal bit cost of encoding pixels can be achieved by minimizing cross-entropy:
\begin{equation}
    E(\mathbf{V}) = \mathbb{E}_{\mathbf{V} \sim P}[-\sum_{t}^{T}log_2{\hat{P}(S_t)}] 
    = \mathbb{E}_{\mathbf{V} \sim P}[-\sum_{t}^{T}\sum_{i,j}^{H,W}log_2{\hat{p}(x_{i,j}, \mathcal{P}({\mathbf{x}_{i,j}, \mathbf{h}_{t-1}}))}].  
\label{eq:loss}
\end{equation}

Eventually, we obtain the optimal probability estimation model.
Each pixel is encoded with an estimation probability distribution $\hat{p}(x_{i,j})$ using the Arithmetic Coding \citep{witten1987arithmetic}.

\begin{figure*}
    \centering
    \includegraphics[width=1.0\textwidth]{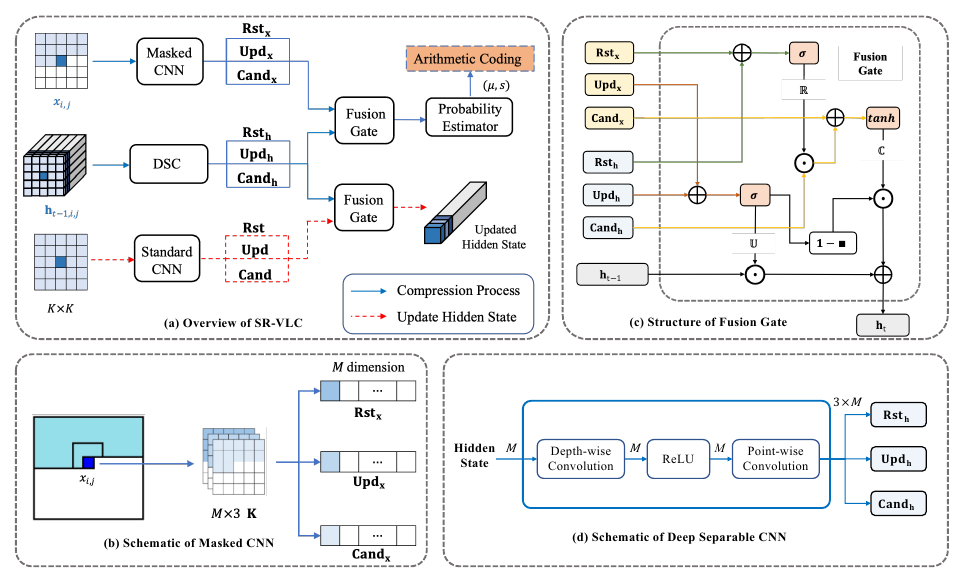}
    \centering
    \caption{An overview of the Streaming Recurrent Lossless Volumetric Compression (SR-LVC), comprising three distinct convolution blocks for extracting multiple features within a slice, two fusion gates for fusing latent features and hidden states, and ultimately obtaining the estimated logistic distribution and updated hidden states.}
    \label{fig:overview}
\end{figure*}

\subsection{Overview}
\figref{fig:overview} illustrates the framework and details of the proposed SR-LVC, encompassing three distinct convolution modules, two fusion gates, and a probability estimator.

When compressing the current pixel $x_{i,j}$, the previous neighboring pixels are first convolved using a Masked-CNN.
As shown in \figref{fig:overview}(b), three $M$-dimensional feature vectors will be extracted when using $M \times 3$ convolution kernel, namely Reset Feature($\mathbf{Rst}$), Update Feature($\mathbf{Upd}$) and Candidate Feature($\mathbf{Cand}$):
\begin{equation}
    \mathbf{Rst}_\mathbf{x}, \mathbf{Upd}_\mathbf{x}, \mathbf{Cand}_\mathbf{x} = \mathcal{MC}(x_{i-K,j-K}, \ldots, x_{i,j-1}; \mathbf{K}).
\end{equation}

Simultaneously, each pixel corresponds to an $M$-dimensional Hidden State $\mathbf{h}_{t-1, i, j}$, encompassing information from previous slices.
The hidden states are also convolved to obtain three feature vectors with the same dimension.
Given that the hidden states $\mathbf{h}_{t-1}$ exist before compressing the current pixel, Masked-CNN is unnecessary.
To minimize the model parameters, convolution operations on hidden states will employ Deep Separable Convolution (DSC) \citep{chollet2017xception}.
As shown in \figref{fig:overview}(d), DSC initially conducts depth-wise convolution on the hidden state to obtain intermediate results with the same dimension.
Subsequently, following the ReLU activation function, point-wise convolution integrates the information from the input channel:
\begin{equation}
    \mathbf{Rst}_\mathbf{h}, \mathbf{Upd}_\mathbf{h}, \mathbf{Cand}_\mathbf{h} = \mathcal{DSC}(\mathbf{h}_{t-1}; \mathbf{K}_{depth}; \mathbf{K}_{point}).
\end{equation}
With the DSC, the number of weights has been reduced from $K^2 \times M^2 \times 3$ to $K^2 \times M + M^2 \times 3$.

After convolution, the resulting $M$-dimensional feature vectors are fed into the fusion gate unit.
Taking inspiration from the GRU, we employ various gating mechanisms to incorporate hidden states containing information from the previous slice into the prediction of the current pixel.
In contrast to the gate mechanism proposed in ICEC, the fusion gate does not contain any parameters and is more suitable for hardware implementation.
As shown in \figref{fig:overview}(c):
\begin{align}
    &\mathbb{R} = \sigma(\mathbf{Rst}_\mathbf{x} + \mathbf{Rst}_\mathbf{h}), \\
    &\mathbb{U} = \sigma(\mathbf{Upd}_\mathbf{x} + \mathbf{Upd}_\mathbf{h}), \\
    &\mathbb{C} = tanh(\mathbf{Cand}_\mathbf{x} + \mathbb{R} \odot \mathbf{Cand}_\mathbf{h}), \\
    &\mathbf{h}_{t,i,j} = \mathbb{U} \odot \mathbf{h}_{t-1, i, j} + (1-\mathbb{U}) \odot \mathbb{C}.
\end{align}
The $\sigma$ and $tanh$ are nonlinear activation functions, respectively, representing the sigmoid and hyperbolic tangent functions.
The $\odot$ denotes the Hadamard product.
The Reset Gate ($\mathbb{R}$) functions in determining which prior information to "forget" \citep{chung2014empirical}.
Multiplying $\mathbb{R}$ with $\mathbf{Cand}_\mathbf{h}$ regulates the impact of the previous slice's state on the current state.
The Update Gate ($\mathbb{U}$) governs the extent to which the hidden state from the previous slice should be "retained" \citep{chung2014empirical}.
These gating mechanisms enable the fusion gate ($\mathcal{FG}$) to capture dependencies between slices effectively.

Subsequently, the new state, incorporating information from both the current and preceding slices, is fed into the probability estimator ($\mathcal{P}$). This estimator utilizes a straightforward linear predictor to obtain the logistic distribution's $(\mu, s)$ values.
Arithmetic Coding employs the discrete logistic distribution to encode the current pixel.
In summary, the formula for the prediction process is as follows:
\begin{equation}
    S_t(\mu), S_t(s) = \mathcal{P}(\mathcal{FG}(\mathcal{MC}(S_t); \mathcal{DSC}(\mathbf{h}_{t-1}); \mathbf{h}_{t-1})).
\end{equation}

Once all neighboring pixels are encoded, updates to hidden states commence after a specific delay.
Illustrated in \figref{fig:overview}(a), the update process mirrors the prediction process, with the distinction that, given the completion of encoding, the hidden state can be updated with all neighboring pixels:
\begin{equation}
    \mathbf{h}_t = \mathcal{FG}(\mathcal{SC}(S_t); \mathcal{DSC}(\mathbf{h}_{t-1}); \mathbf{h}_{t-1}).
\end{equation}
Finally, transmits the updated hidden states to the next slice, facilitating the transfer of context information.

\subsection{Variable-scale Discrete Logistic Function}
\label{sec:logistic}
The value of the model weights can be determined by minimizing the cross-entropy $E(\mathbf{V})$:
\begin{equation}
    E(\mathbf{V}) = -\sum_{t}^{T}log_2{\hat{p}(S_t; S_t(\mu); S_t(s))}.
\end{equation}
In practice, discrete logistic likelihood functions $\hat{p}$ are often used to estimate the exact probability distribution:
\begin{equation}
    \hat{p}(x, \mu, s) = CDF_{\hat{p}}(x + 0.5, \mu, s) -  CDF_{\hat{p}}(x - 0.5, \mu, s).
\end{equation}

However, in scenarios with high sampling bits, learning-based methods face the issue of gradient vanishing, leading to sub-optimal models.
To alleviate this issue, we introduce a scaling factor $L$ during normalization to further adjust the logistic distribution following the model output $(\mu, s)$.
When the bit depth is $D$, its normalization and probability formula are:
\begin{align}
    &x_n = \mathcal{N}(x) = (x / 2^D) \times L, \quad
     b = \mathcal{N}(0.5), \\
    \hat{p}(x) = &CDF_{\hat{p}}(x_n + b, \mu_n, s) - CDF_{\hat{p}}(x_n - b, \mu_n, s).
\end{align}

\begin{figure}
    \includegraphics[width=11 cm]{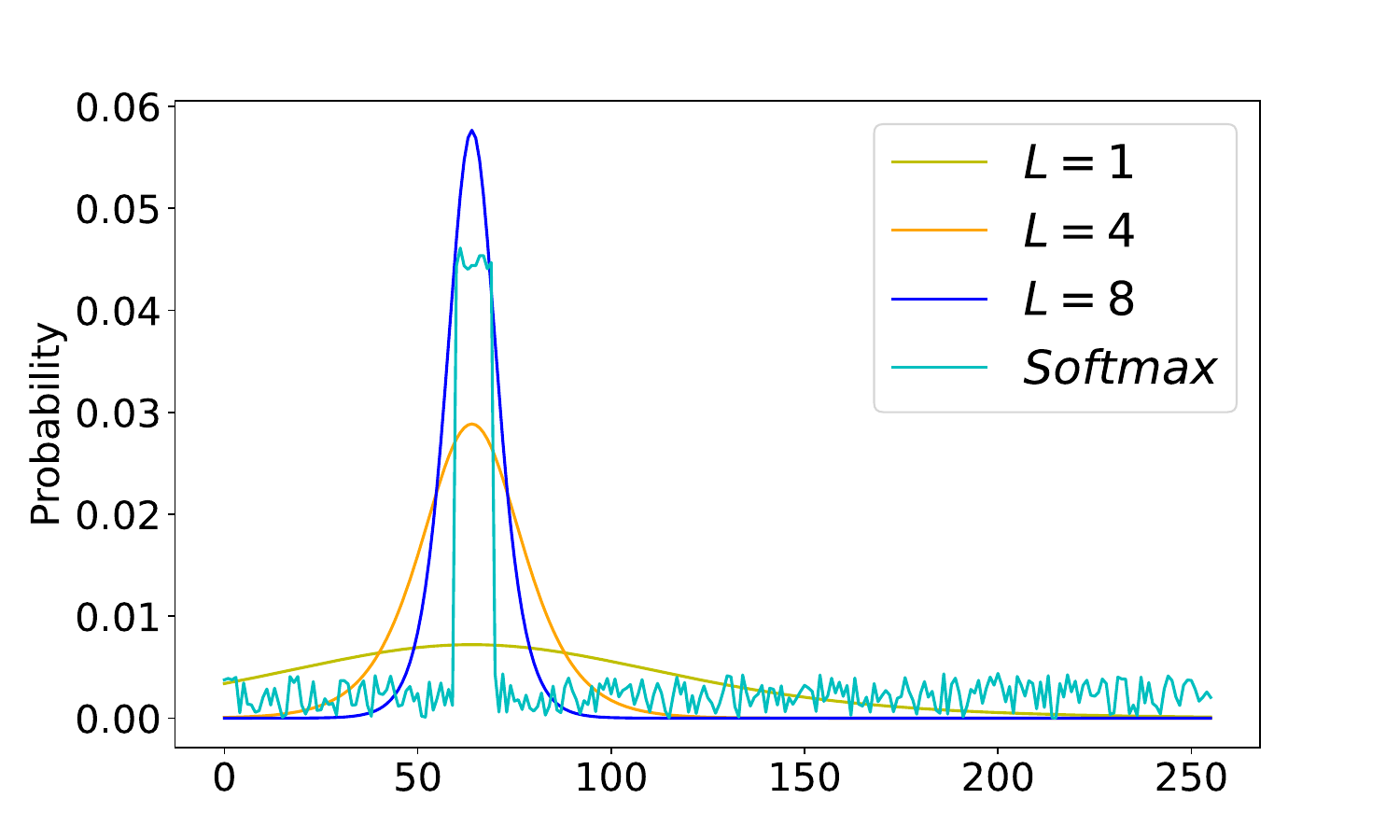}
    \centering
    \caption{Comparison between the different scaling factor $L$ logistic likelihood functions when $D=8;\mu=64;s=e^{-2}$.}
    \label{fig:log}
\end{figure}

\figref{fig:log} compares the probability distributions of the softmax function and different scaling factor $L$ logistic likelihood functions when $D=8$.
Upon determining the output of the probability estimator, a larger $L$ can concentrate the probability distribution more closely around the mean $\mu$.
Employing a larger scaling factor $L$ renders the distribution more similar to the output of the softmax function, fostering advantages in model training and enhancing the model's ability to fit the actual data distribution.
More experimental verification will be provided in Section \ref{sec:ablation} (Ablation Study).

\subsection{Encoding and Decoding}
In Arithmetic Coding, the predicted probability $\hat{p}(x_{i,j})$ is employed to allocate the corresponding number of bits to the current pixel $x_{i,j}$.
However, utilizing the variable-scale discrete logistic likelihood function results in a few pixels with predicted probabilities lower than the precision of arithmetic coding, making them impossible to encode.
Hence, we preserve the positions and values of pixels that cannot be accurately compressed by arithmetic coding for recovery during decompression.

The decompression process closely mirrors the compression process.
It employs hidden states $\mathbf{h}_{t-1}$ and decompressed pixels to estimate the probability distribution of the pixel to be decompressed.
The decoder retrieves the pixel value from the compressed bitstream using the predicted probability distribution.
% =============================================================

% =============================================================
\section{Hardware Implementation}
Deep learning has demonstrated remarkable capabilities in addressing lossless compression.
However, it is imperative to consider Data Access Time when enhancing the compression ratio.
In this paper, leveraging the streaming nature of our lightweight framework, we implemented SR-LVC on an FPGA.
To the best of our knowledge, this represents the inaugural implementation of a learning-based accelerator for lossless volumetric compression on an FPGA.

\begin{figure*}
    \centering
    \includegraphics[width=1.0\textwidth]{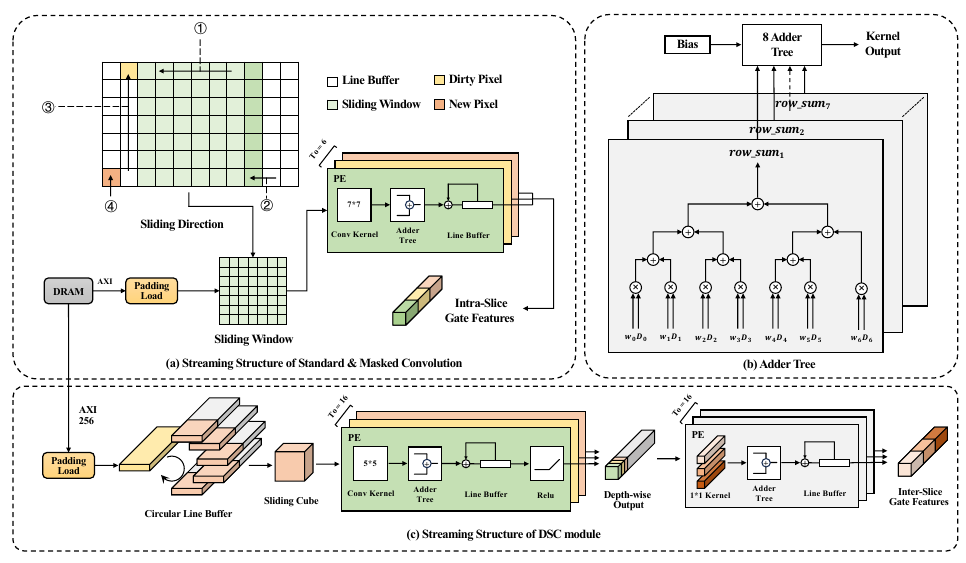}
    \centering
    \caption{Hardware Implementation Overview of the Streaming Recurrent Lossless Volumetric Compression (SR-LVC).}
    \label{fig:stream}
\end{figure*}

\subsection{Streaming Design for Convolution Module}
The architecture proposed in this paper primarily comprises three main convolution modules: Masked CNN, Standard CNN, and Depth-wise Separable Convolution (DSC) module.
Since the distinction between Masked CNN and Standard CNN is solely in the weight placement within the convolution kernel, in this chapter, we use Standard CNN and the DSC module as examples to offer a detailed description of the streaming structure of these convolution modules.

\subsubsection{Standard CNN}
A standard CNN module comprises a sliding buffer and a convolution accelerator.
The sliding buffer utilizes Block RAM (BRAM) to store pixel information retrieved from Dynamic RAM (DRAM) and continually updates the sliding window registers for real-time calculations.
For a convolution kernel size of $K \times K$, processing a gray image of width $W$ necessitates $K$ linear buffers, each with a length of $K+W-1$.
Simultaneously, a sliding window moves from left to right along the line buffer with a stride of $1$, placing the image pixels within a sliding window into the convolution accelerator.
As shown in \figref{fig:stream}(a), the sliding window's movement involves four steps:
\begin{itemize}
    \item[1] {All values stored in the registers shift leftwards by one unit, implying the data in the leftmost column will be discarded.}
    \item[2] {Newly adjacent pixel values are read into the rightmost column of the sliding window, equivalent to sliding the sliding window one step to the right.}
    \item[3] {The left-adjacent column of the sliding window shifts upwards by one unit, removing the topmost 'dirty' pixel.} 
    \item[4] {A new pixel value is read from DRAM to the bottom of that column, completing the update for the linear buffer.}
\end{itemize}

The Processing Elements (PE) accomplish convolution acceleration on the sliding window by fully unfolding the multiply-add operations.
Each PE incorporates an Adder Tree to finalize the convolution acceleration.
Illustrated in \figref{fig:stream}(b), considering a $7 \times 7$ convolution kernel as an example, we initially employ $7$ input multiply-add trees to compute each row of the convolution.
Subsequently, an additional addition tree aggregates the results of each row and adds the bias.
Additionally, we instantiate $6$ convolution acceleration PEs, enabling the concurrent computation of up to $6$ convolution kernels ($To=6$).

The masked convolution kernel has only half the weight compared to the standard convolution.
Consequently, it is necessary to adjust the sliding window and PEs to obtain the final multiple gate features.

\subsubsection{Depth-wise Separable Convolution}
For the $M$-dimensional hidden states, using the same structure as the linear buffer of the standard CNN module would require shifting each read value across multiple channels, leading to a more substantial buffer update time.
Consequently, we designed a cyclic on-chip cache structure for the line buffer in the DSC module.
As shown in \figref{fig:stream}(c), when the size of the convolution kernel is $K \times K$, the number of linear buffers is $K + 1$, the width of each linear buffer is $M$, and the length is also $K + W - 1$.
At each moment, only one linear buffer is invalid, and the AXI engine reads the hidden states from the DRAM.
Meanwhile, the slider data of size $K \times K \times M$ is read sequentially from the other $K$ valid linear buffers and passed to the DSC acceleration module for calculation.

Leveraging the characteristics of deep-wise separable convolution, we utilize a parallel accelerator with $To=M$ parallelism for the DSC module, independently processing each channel.
For each channel, the multiply-add results of each row of the convolution kernel are computed in parallel.
Subsequently, the results of each row are aggregated using the addition tree to obtain the $M$-dimensional convolution result.
Similarly, set the parallelism of the point-wise convolution layer to $To=M$.
The DSC convolution yields the gate features after passing through the ReLU activation and point-wise convolution layers.

\subsection{Implementation Details}
\subsubsection{Design of Fusion Gate}
After the operation of each convolution module, multiple $M$-dimensional vectors are obtained, corresponding to Reset features, Update features, and Candidate features for the fusion gate.
The fusion gate comprises multiple adders, multipliers, and nonlinear activation units.
For better compatibility with hardware implementation, we employ the hard sigmoid function and hard tanh activation function during both training and deployment.
\begin{align}
    \sigma_{hard}(x) &= 
    \begin{cases}
        0, & \text{if } x \leq -3 \\
        1, & \text{if } x \geq 3 \\
        x \div 6 + 0.5, & \text{otherwise } \\
    \end{cases} \\
    {tanh}_{hard}(x) &= 
    \begin{cases}
        -1, & \text{if } x \leq -1 \\
        1, & \text{if } x \geq 1 \\
        x, & \text{otherwise } \\
    \end{cases}
\end{align}
The hard activation function has a more straightforward form, can be easier to implement on hardware, and is still able to capture the nonlinear features required by the gate unit \citep{hubara2016binarized}.

\subsubsection{Design of Probability Estimator}
The goal of the probability estimator is to produce logistic distribution parameters $(\mu, s)$.
Consequently, we utilize $M \times 2$ DSPs to fully parallelize the multiplication processing and two adder trees to calculate the results.
Ultimately, the probability results and the updated hidden states will be written back to the DRAM.

\subsubsection{Deployment}
\label{sec:deploy}
The convolution layers, fusion gates, and probability estimators are processed in a pipeline.
Following a certain start-up delay, the system can streamwise output the probability results.
Arithmetic Coding operates on the CPU, reading the $(\mu, s)$ from DRAM, encoding the pixel value, and storing it on the hard disk.
Simultaneously, when deploying the trained model to the hardware platform, the obtained model weights need to be truncated and expressed as $16$-bit half-precision floating-point numbers to optimize resource utilization.

% =============================================================

\section{Experimental Results}
This section compares and evaluates the Streaming Recurrent Lossless Volumetric Compression system (SR-LVC) proposed in this paper with the traditional 3-D image compression algorithm and other compression algorithms based on deep learning models on various 3-D medical datasets.

\subsection{Experiment Protocol}
\subsubsection{Training Details}

% =============================================================
\renewcommand{\arraystretch}{1.4}
\begin{table}[h]
\caption{Hyperparameters of SR-LVC}
\label{tab:hyper}
    \begin{center}
        \begin{tabular}{@{}cc|cc@{}}
        \hline
        \multicolumn{2}{c|}{\textbf{Model Hyperparameters}} & \multicolumn{2}{c}{\textbf{Training Hyperparameters}}                                           \\ 
        \hline
        Kernel Size                        & $7 \times 7$\tnote{1}               & Epoch                 & 1000                                                                     \\
        Feature Dimension $M$            & 16             & Initial Learning Rate & 0.0005                                                                   \\
        Convolution Bias             & True           & Updated Stride          & 5                                                                        \\
        Activation Function          & Hard                & Optimizer             & Adam                                                                     \\
                                           &                & Scaling Factor $L$       & \begin{tabular}[c]{@{}c@{}}1 (8 bit depth) \\ 8 (12 bit depth) \end{tabular} \\ 
        \hline           
        \end{tabular}
        \begin{tablenotes}
            \footnotesize
            \item[1]  In the DSC module, the kernel size is adjusted to $5\times5$.
        \end{tablenotes}
    \end{center}
\end{table}
% =============================================================

\tabref{tab:hyper} lists all the hyperparameters of the SR-LVC. 
For a 3-D gray image input, the number of parameters is only $4866$ according to the specified hyperparameters.
Typically, the training method for recurrent neural networks involves the Back Propagation Through Time (BPTT) algorithm \citep{werbos1990backpropagation}.
However, when dealing with long-time sequences, the issue of vanishing gradients \citep{graves2012long} arises, corresponding to the large number of slices in our volumetric image compression.
To address this issue, we introduce a parameter known as the "Updated Stride," set to $5$.
With this parameter, we update only the gradients within $5$ consecutive slices, avoiding the propagation of gradients over the entire long sequences during the backward propagation process.
This strategy effectively mitigates the vanishing gradient problem associated with long sequences, ensuring the transmission of hidden states and enhancing the stability of the training process.
Additionally, for scenarios with a bit depth of $12$ or more, we set the Scaling Factor $L$ to $8$.

\subsubsection{Metrics}
To evaluate SR-LVC on the validation dataset and compare it with traditional baseline algorithms and state-of-the-art learning-based frameworks, we use Bits per Pixel (BPP) as an Evaluation Metric (EM).
The formula is $C(\mathbf{V})/(T \times W \times H)$ \cite{chen2022exploiting}, where $T$ is the number of slices in the volumetric image $\mathbf{V}$, and $W$ and $H$ are the width and height of each slice.
$C(\mathbf{V})$ is the total size of the compressed volume $\mathbf{V}$ after Arithmetic Coding. 
It includes compressed file headers and the locations and values for pixels that cannot be compressed using Arithmetic Coding.

\subsubsection{Baseline Methods}
We compare our proposed approach with traditional 2-D image compressors like PNG \citep{roelofs1999png} and JPEG2000 \citep{taubman2002jpeg2000}.
The traditional 3-D video-based compression algorithms, including JP3D \citep{bruylants2009jp3d}, H.265/HEVC \citep{sullivan2012overview}, and FFV1 \citep{niedermayer2013ffv1}, are also selected as the baseline methods.
For PNG compression, we use the Python Imaging Library (PIL). 
For JPEG2000 and JP3D, we use OpenJPEG-2.3.1 from the OpenJPEG library. 
The FFmpeg-4.3 is used to compress images with HEVC and FFV1. 
All compression methods except HEVC adopt the default setting for lossless compression. 
To achieve the optimal compression ratio, we set the configuration of HEVC to "very slow". 

For learning-based lossless image compression, we compared our method with L3C \citep{mentzer2019practical}, ICEC \citep{chen2022exploiting}, and aiWave \citep{xue2022aiwave}.
According to the \tabref{tab:compare}, L3C cannot compress datasets with a bit depth of more than $8$ bits, and ICEC and aiWave cannot directly compress datasets with varying image resolutions.

\subsubsection{Platform}
During the training phase, we use the NVIDIA RTX 3090 GPU. 
For CPU runtime, the Intel(R) Xeon(R) Gold 6128 CPU running at 3.4 GHz is employed. 
Finally, our SR-LVC is deployed on the Xilinx Alveo U250 using High-Level Synthesis (HLS) techniques for further evaluation.

\subsection{Datasets}
Nine 3-D medical image datasets, encompassing CT and MRI scans, were utilized for extensive evaluation. 
Similar to other learning-based compression systems, distinct sets of neural network parameters were trained on individual datasets, and the compression ratio was validated on the test dataset. 
Details of the 3-D medical datasets are provided in \tabref{tab:dataset}.

% =============================================================
\renewcommand{\arraystretch}{1.5}
\begin{table*}
\caption{Basic information of 3-D medical datasets}
\label{tab:dataset}
    \begin{center}
        \begin{tabular}{@{}ccccc@{}}
        \hline
        \textbf{Name}               & \textbf{Type}  & \textbf{\begin{tabular}[c]{@{}c@{}}Bit Depth\\ (Bits)\end{tabular}} & \textbf{\begin{tabular}[c]{@{}c@{}}Resolution\\ (Pixels)\end{tabular}} & \textbf{\begin{tabular}[c]{@{}c@{}}Training/Validation Set\\ Number\end{tabular}} \\ 
        \hline
        \multirow{3}{*}{MRNet}      & MRI (Axial)    & \multirow{3}{*}{$8$}                                                  & \multirow{3}{*}{$(17\sim61)\times256\times256$}                                         & \multirow{3}{*}{$1130/120$}                                                             \\
                                    & MRI (Coronal)  &                                                                     &                                                                       &                                                                                       \\
                                    & MRI (Sagittal) &                                                                     &                                                                       &                                                                                       \\
        \hline
        \multirow{3}{*}{CHAOS}      & CT             & \multirow{3}{*}{$12$}                                                 & $(78\sim294)\times512\times512$                                                         & \multirow{3}{*}{$28/12$}                                                                \\
                                    & MRI (T2SPIR)   &                                                                     & $(26\times256\times256)\sim(44\times512\times512)$                                                               &                                                                                       \\
                                    & MRI (T1DUAL)   &                                                                     & $(26\times256\times256)\sim(50\times400\times400)$                                                        &                                                                                       \\
        \hline
        \multirow{3}{*}{DeepLesion} & CT Study 01    & \multirow{3}{*}{$12$}                                                 & $(6\sim260)\times(512\times512\sim768\times768)$                                                               & $1531/382$                                                                              \\
                                    & CT Study 02    &                                                                     & $(6\sim400)\times(512\times512\sim768\times768)$                                                              & $1268/316$                                                                              \\
                                    & CT Study 03    &                                                                     & $(6\sim260)\times(512\times512\sim768\times768)$                                                                & $932/233$                                                                               \\ 
        \hline
        \end{tabular}
    \end{center}
\end{table*}
% =============================================================

\subsubsection{MRNet}
This extensive knee MRI dataset encompasses 1370 knee examinations conducted on 1199 patients \citep{bien2018deep}.
Each examination includes three distinct scanning orientations: axial, coronal, and sagittal.
Within each scanning orientation, a sequence of image slices, ranging from $17$ to $61$ slices, is provided, with a resolution of $256\times256$ pixels for each image slice, all stored with an $8$ bit depth. 
The dataset has been meticulously partitioned into two subsets: a training set, comprising 1130 scans derived from 1088 patients, and a validation set, comprising 120 scans sourced from 113 patients. 
We directly utilize the established training and validation divisions in our experiment.

\subsubsection{CHAOS}
The Chaos dataset \citep{kavur2021chaos} comprises abdominal CT and MRI images from 40 patients.
In our experiment, we selected the entire CT images, the out-phase channel of T1DUAL MRI images, and the T2SPIR MRI images for lossless compression.
All images were initially stored in $16$-bit DICOM format, yet the actual sampling depth was $12$-bit.
We initially arranged the 40 patients in ascending order by index.
Subsequently, we designated the first $70\%$ of the images (28 patients) as the training set, while the remaining $30\%$ of images (12 patients) constituted the validation set.

\subsubsection{DeepLesion}
The DeepLesion dataset \citep{yan2018deeplesion} stands as one of the largest open clinical medical CT image datasets to date, comprising data from 4427 anonymous patients, amounting to 10594 CT scans (averaging approximately 3 follow-up scans per patient). 
Each image slice is named in the format "\{Patient Index\}\_\{Study Index\}\_\{Series Index\}". 
The dataset consists of a total of 26 distinct studies.
Although the CT images in the dataset are stored in an unsigned 16-bit format, to obtain the original Hounsfield unit values, it is necessary to subtract 32768 from pixel values to get a slice image with a window width size of [-1024, 3071].
Given the extensive size of the DeepLesion dataset, occupying a substantial 221GB of storage space and distributing across 56 zip files, we select the CT images from studies 01, 02, and 03 contained within the first 20 zip files. 
Subsequently, we sort these images by patient index and partition them based on consecutive slice indices. 
Then, we use the top $80\%$ of CT images from each study as the training set, while the remaining $20\%$ of images are allocated to the validation set. 

\subsection{Lossless Compression Results}

% =============================================================
\renewcommand{\arraystretch}{1.5}
\begin{table*}
\caption{Lossless Compression Results (In Average BPP) of SR-LVC and Other Lossless Compression Methods on Various Datasets}
\label{tab:result}
    \begin{center}
        \begin{tabular}{cccccccccc}
        \hline
        \textbf{}    & PNG               & JP2K             & JP3D             & HEVC              & FFV1              & L3C               & ICEC               & aiWave             & SR-LVC             \\
                     & \multicolumn{9}{c}{-----------------------------------Bits per Pixel-----------------------------------} \\ \hline
                     & \multicolumn{9}{c}{MRNet}                                                                                                                                                            \\ \cline{2-10} 
        Axial        & 5.327              & 4.993             & 4.966             & 5.228             & 4.865             & 5.189             & 4.642             & 4.545              & \textbf{4.474}              \\
        Coronal      & 4.571              & 4.161             & 4.152             & 4.385             & 4.049             & 4.342             & 3.841             & 3.804              & \textbf{3.707}              \\
        Sagittal     & 5.570              & 5.306             & 5.289             & 5.595             & 5.203             & 5.598             & 4.974             & \textbf{4.829}              & 4.862              \\ \cline{2-10} 
                     & \multicolumn{9}{c}{CHAOS}                                                                                                                                                            \\ \cline{2-10} 
        CT           & 7.472              & 5.227             & 5.287             & 5.245             & 5.303             & -                 & -                 & 5.077              &\textbf{ 4.293}              \\
        MRI (T2SPIR) & 5.380              & 3.758             & 3.881             & 4.070             & 3.746             & -                 & -                 & -                  & \textbf{3.471}              \\
        MRI (T1DUAL) & 5.341              & 3.753             & 3.879             & 4.056             & 3.714             & -                 & -                 & -                  & \textbf{3.312}              \\ \cline{2-10} 
                     & \multicolumn{9}{c}{DeepLesion}                                                                                                                                                       \\ \cline{2-10} 
        CT Study 01  & 7.277              & 5.478             & 5.597             & 5.664             & 5.430             & -                 & -                 & -                  & \textbf{4.664}              \\
        CT Study 02  & 7.150              & 5.365             & 5.485             & 5.445             & 5.316             & -                 & -                 & -                  & \textbf{4.643}              \\
        CT Study 03  & 7.367              & 5.591             & 5.708             & 5.664             & 5.549             & -                 & -                 & -                  & \textbf{4.852}              \\ \hline
        \end{tabular}
    \end{center}
\end{table*}
% =============================================================

\tabref{tab:result} presents the lossless compression results (in average BPP) on various validation datasets.

\subsubsection{MRNet}
On the MRNet Dataset, it is noteworthy that SR-LVC outperforms all 2-D image compressors.
Specifically, our method achieves a significant BPP reduction of over $8\%$ ($10.4\%$ for Axial, $10.9\%$ for Coronal, and $8.4\%$ for Sagittal) compared to JPEG2000.
In comparison to the traditional 3-D image compression method JP3D, our method achieves a BPP reduction exceeding $8\%$ ($9.9\%$ for Axial, $10.7\%$ for Coronal, and $8.1\%$ for Sagittal).
Furthermore, SR-LVC outperforms the state-of-the-art video compressor FFV1, resulting in a BPP reduction exceeding $6\%$ ($8.0\%$ for Axial, $8.4\%$ for Coronal, and $6.6\%$ for Sagittal).

Additionally, when compared with deep learning-based methods, our approach surpasses L3C and ICEC across all MRNet dataset scans.
In comparison to L3C, we attain a notable BPP reduction of more than $13\%$ ($13.8\%$ for Axial, $14.6\%$ for Coronal, and $13.1\%$ for Sagittal).
Similarly, in comparison to ICEC, we achieve a BPP reduction of $3.6\%$, $3.5\%$, and $2.3\%$ for the Axial, Coronal, and Sagittal datasets, respectively.
Furthermore, when compared with aiWave, whose weighting parameters are orders of magnitude greater than ours, the proposed framework achieves better compression ratios in two scans ($1.6\%$ for "Axial" and $2.5\%$ for "Coronal").

\subsubsection{CHAOS}
On the CT dataset, SR-LVC significantly decreases the mean BPP by $17.9\%$ compared to JPEG2000 and $19.0\%$ compared to FFV1.
In the MRI datasets, FFV1 performs the best among the traditional methods, with SR-LVC exhibiting a $7.3\%$ decrease in mean BPP (MRI-T2SPIR) and $10.8\%$ (MRI-T1DUAL).
Due to the CHAOS dataset's sampling bit depth of 12 bits, it cannot be compressed directly using L3C.
Additionally, ICEC and aiWave cannot process datasets directly with varying slice resolutions.
For CHAOS-CT, aiWave selects 1000 cube blocks of $64\times64\times64$ for the compression experiment instead of compressing the entire validation dataset, making its average BPP for reference only.
However, the proposed method yielded a $15.4\%$ BPP (\textit{i.e.}, $4.393$bpp vs. $5.077$bpp) improvement compared to aiWave.

\subsubsection{DeepLesion}
On the DeepLesion dataset, given its substantial data volume and variable slice resolutions, compression can only be performed by SR-LVC and traditional image compressors.
The experimental results demonstrate that SR-LVC outperforms traditional image compressors in all cases.
Compared to JPEG2000, it achieves an average BPP reduction of $14.9\%$, $13.5\%$, and $13.2\%$ in three studies, respectively.
In comparison to JP3D, the average BPP reduction is over $15\%$ ($16.7\%$ for study 01, $15.4\%$ for study 02, and $15.0\%$ for study 03).
In comparison with the video-based compression algorithm FFV1, the average BPP reduction also exceeds $12\%$ ($14.1\%$, $12.7\%$, and $12.6\%$, respectively).

\subsection{Generalization Experiments}
Taking into account the feasibility of practical applications, we conducted several experiments to assess the robustness and model generalization of SR-LVC.
We directly deploy the model trained on one dataset to compress data from a different validation dataset.
Specifically, we conducted three distinct migration experiments, which included:

\subsubsection{Same Dataset, Similar Scan Mode}

\begin{figure*}
    \includegraphics[width=16cm]{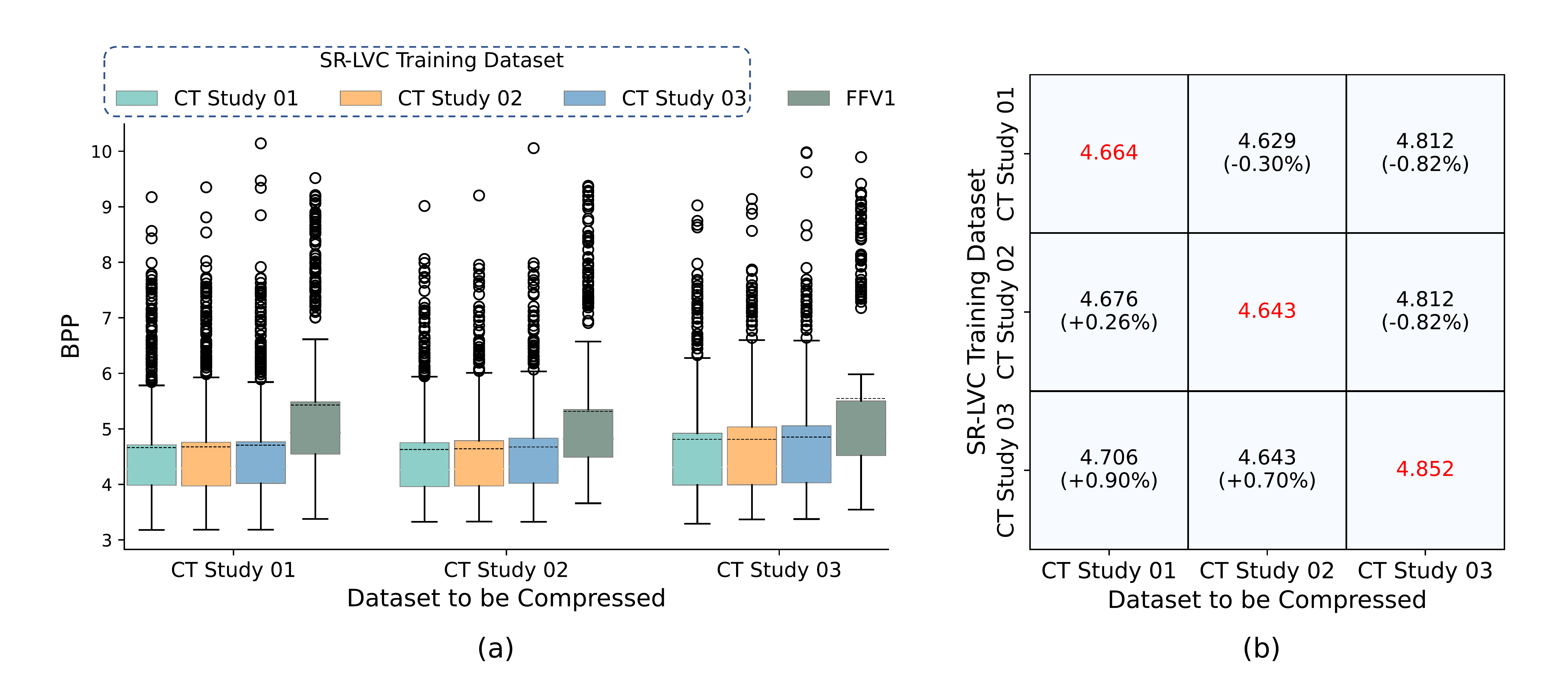}
    \centering
    \caption{BPP Distributions and Average BPP Results on The DeepLesion Datasets Using Different Training Datasets.}
    \label{fig:gen_res2}
\end{figure*}

The categories of DeepLesion datasets exhibit relatively similar image features.
\figref{fig:gen_res2}(a) illustrates the BPP distribution for each category, with the figure's legend indicating the dataset on which the model was trained and validated, and subsequently used for compression on each class of the validation dataset.
\figref{fig:gen_res2}(b) presents the average BPP results, where the columns denote the SR-LVC training and validation dataset, and the rows represent the test dataset to be compressed.

The experimental results demonstrate remarkable consistency in performance when transferring models among scanning modes with similar characteristics.
The average BPP for all categories is superior to that of FFV1 and remains stable.
However, in CT study 03, the training data is the most limited among the three categories, resulting in the model trained on it performing less effectively than the others.

\subsubsection{Same Dataset, Different Scan Mode}

\begin{figure*}
    \includegraphics[width=16cm]{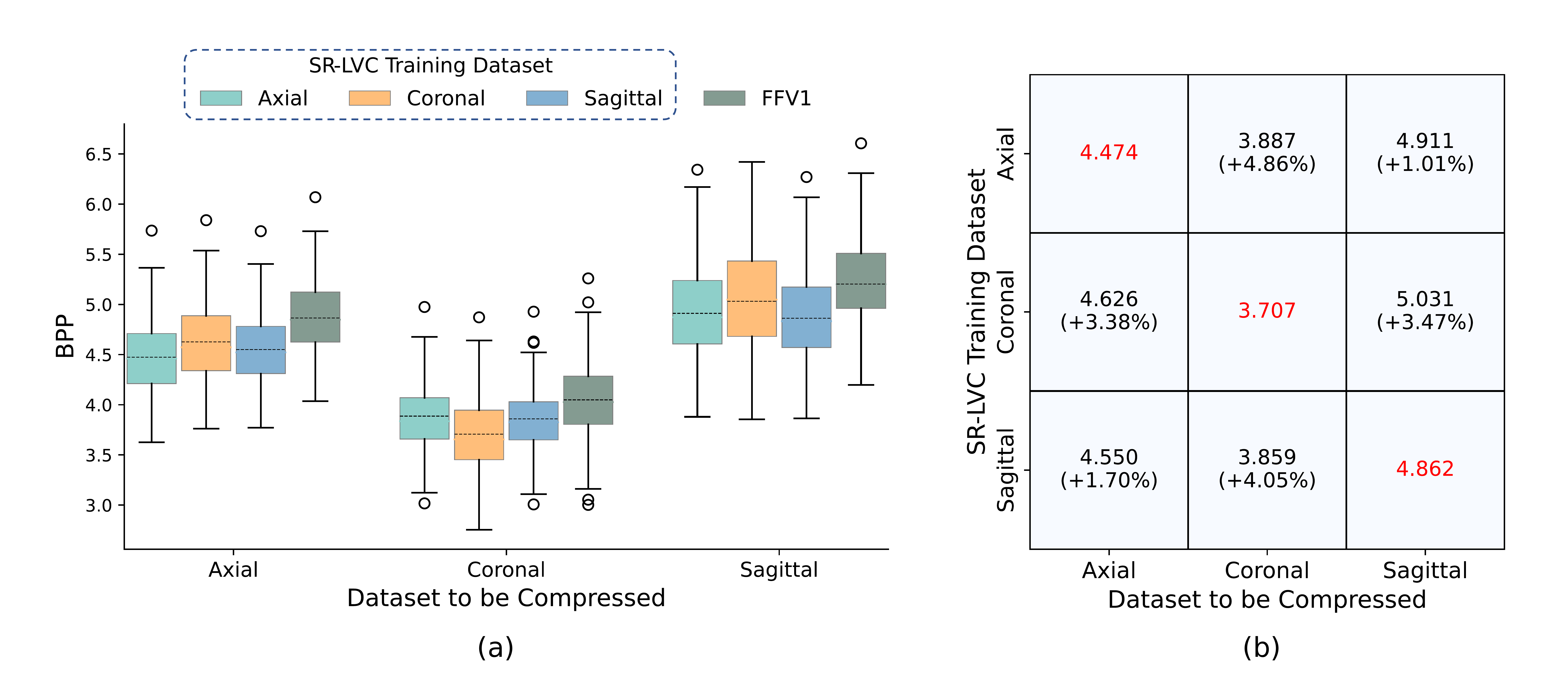}
    \centering
    \caption{BPP Distributions and Average BPP Results on The MRNet Datasets Using Different Training Datasets.}
    \label{fig:gen_res1}
\end{figure*}

In the MRNet dataset, the Axial, Coronal, and Sagittal categories employ different scanning methods, and they display significant morphological differences, posing a challenge to the model's performance in handling diverse anatomical structures and image morphologies.
\figref{fig:gen_res1} illustrates the results of the generalization experiments on the MRNet dataset.
Specifically, the maximum observed performance decrease on the three validation datasets for different categories is $3.4\%$, $4.9\%$, and $3.5\%$, respectively.
However, when compared with the video-based method FFV1, it still maintains competitive compression performance.
This suggests that models trained on one category's dataset can be easily applied to other categories with different scan modes, with a limited decrease in compression performance.

\subsubsection{Different Dataset, Similar Scan Mode}

\begin{figure}
    \includegraphics[width=10 cm]{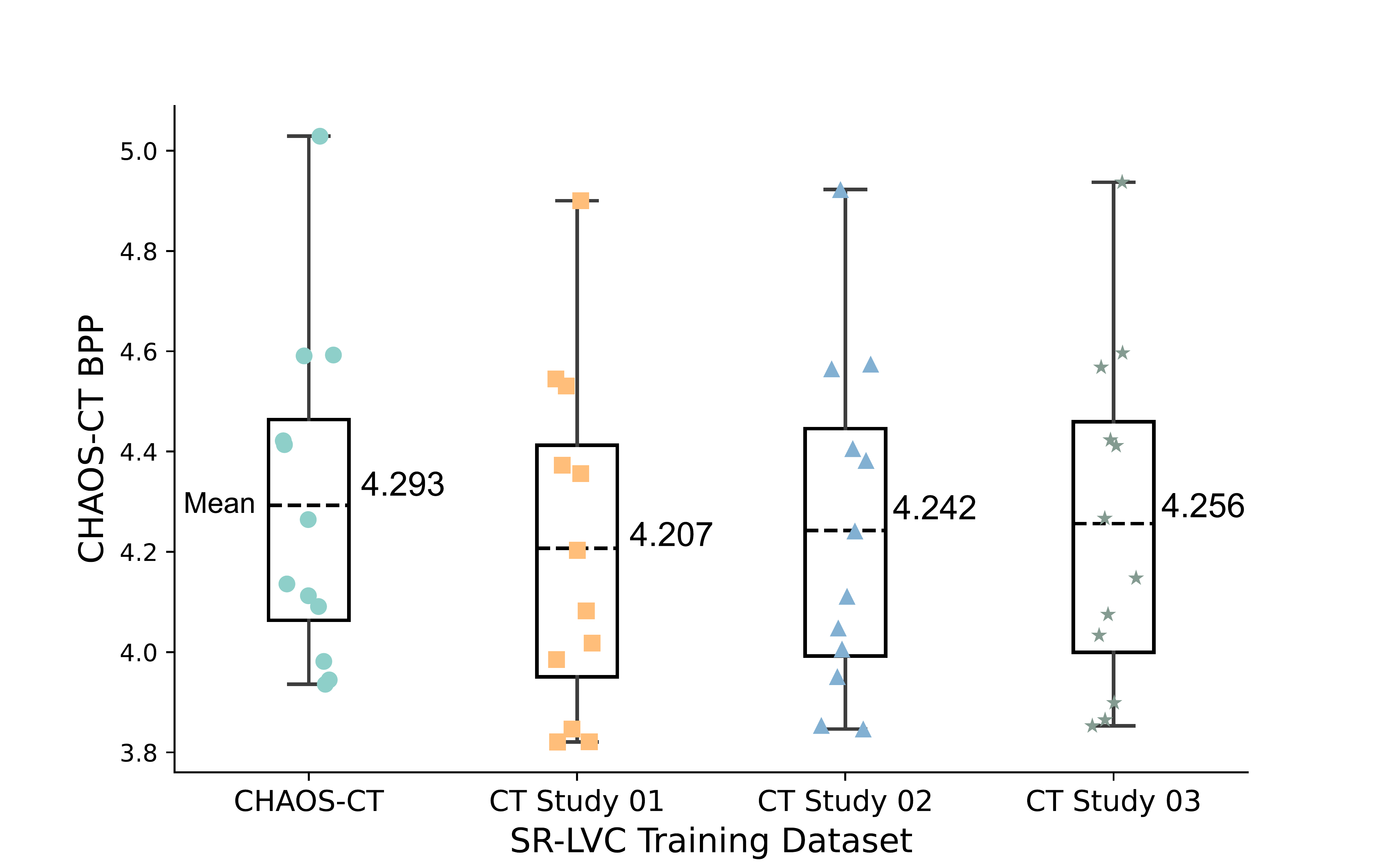}
    \centering
    \caption{BPP Distributions on the CHAOS-CT Dataset Using Different Training Datasets.}
    \label{fig:gen_res3}
\end{figure}

To conduct generalization experiments across different datasets, we employ models trained on the large-scale DeepLesion dataset to compress the CHAOS-CT dataset, validating SR-LVC's adaptability and generalization across diverse datasets.
The experimental results in \figref{fig:gen_res3} demonstrate that models trained on the DeepLesion dataset can successfully compress data from other datasets.
Furthermore, as these models are trained on large-scale datasets, their compression performance on CHAOS-CT is even superior to the model trained on CHAOS-CT itself.
The maximum reduction in average BPP reaches up to $2.0\%$.
This demonstrates that our model can be transferred across different datasets with similar scanning methods.

\subsection{Ablation Study}

% =============================================================

\subsubsection{Analysis of Hidden States Extracting}

\begin{figure}[ht]
    \includegraphics[width=10 cm]{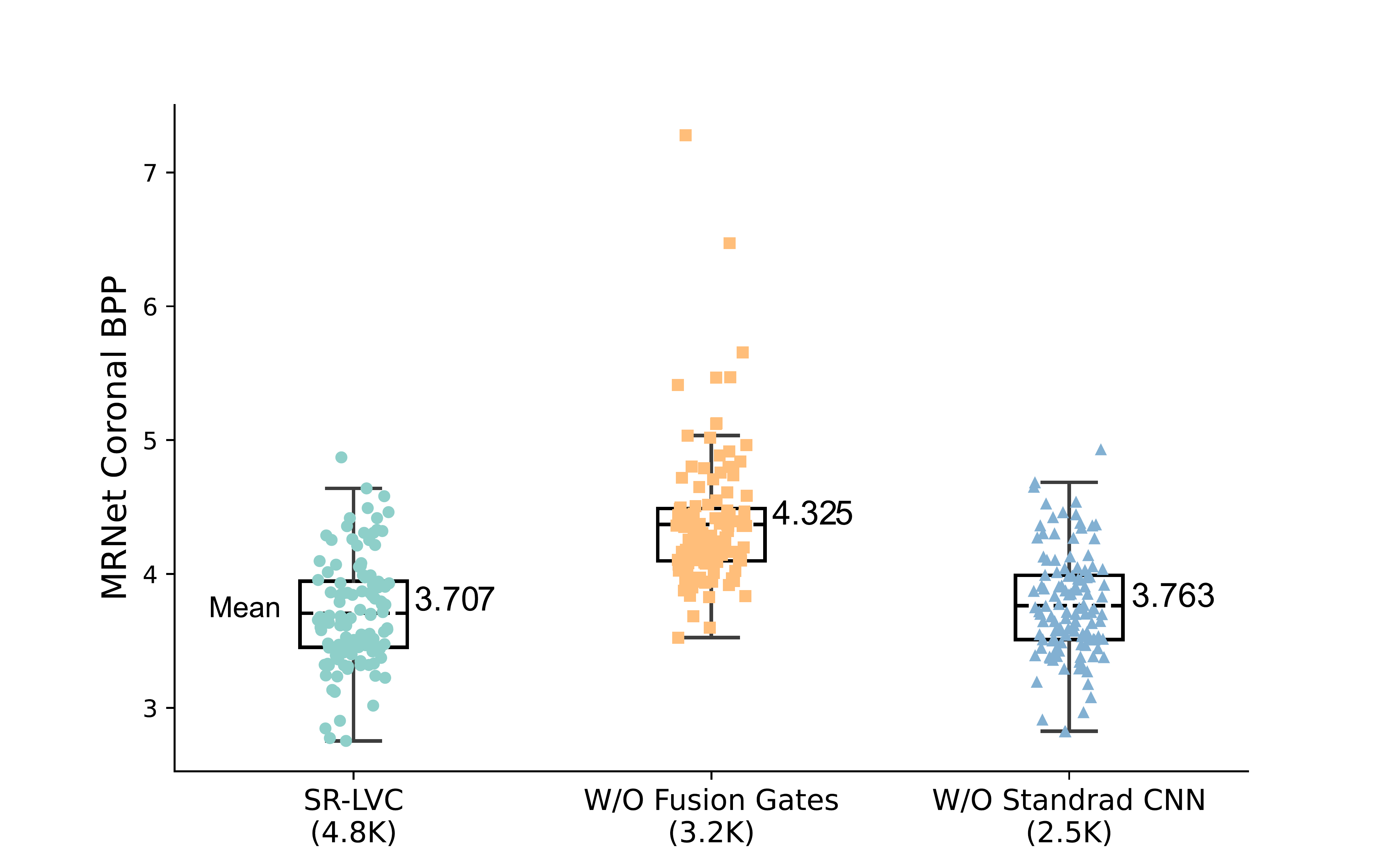}
    \centering
    \caption{BPP Distributions on The MRNet-Coronal Dataset Using Different Alternative Variants.}
    \label{fig:ablation}
\end{figure}

In SR-LVC, the hidden states retain the information from previous slices.
To assess the impact of inter-slice information on the overall lossless compression performance, we eliminate the fusion gates and devise an alternative variant of our method.
To ensure fairness in the experimental comparison after removing the fusion gate units, we adjusted the feature dimension $M$ to $128$.
With this modification, the model's parameters amount to $3.2$K.
On the other hand, in SR-LVC, we applied a standard convolution layer to the current slice for updating the hidden states.
To assess the updating process, we devise an alternative variant of our method by removing the standard CNN and directly using the states of the input probability estimator as the updated hidden states.

We subsequently conducted the retraining on the MRNet-Coronal dataset, following the same training procedures outlined in Section V (Experiment Protocol).
\figref{fig:ablation} illustrates that the variant, which removes all the fusion gates, experiences a substantial degradation in compression ratio ($16.7\%$ in BPP compared to SR-LVC).
Without the standard CNN, its average BPP was slightly inferior to that of SR-LVC.
The standard CNN possesses a larger receptive field, which is advantageous for extracting intra-slice information.
This proves that the contribution of inter-slice dependencies to the compression ratio is greater than that of intra-slice dependencies.
Additionally, it also highlights the flexibility of SR-LVC, which can accomplish the compression task with fewer parameters without compromising the compression ratio.

\subsubsection{Analysis of Scaling Factor}
\label{sec:ablation}
\begin{figure}
    \includegraphics[width=9 cm]{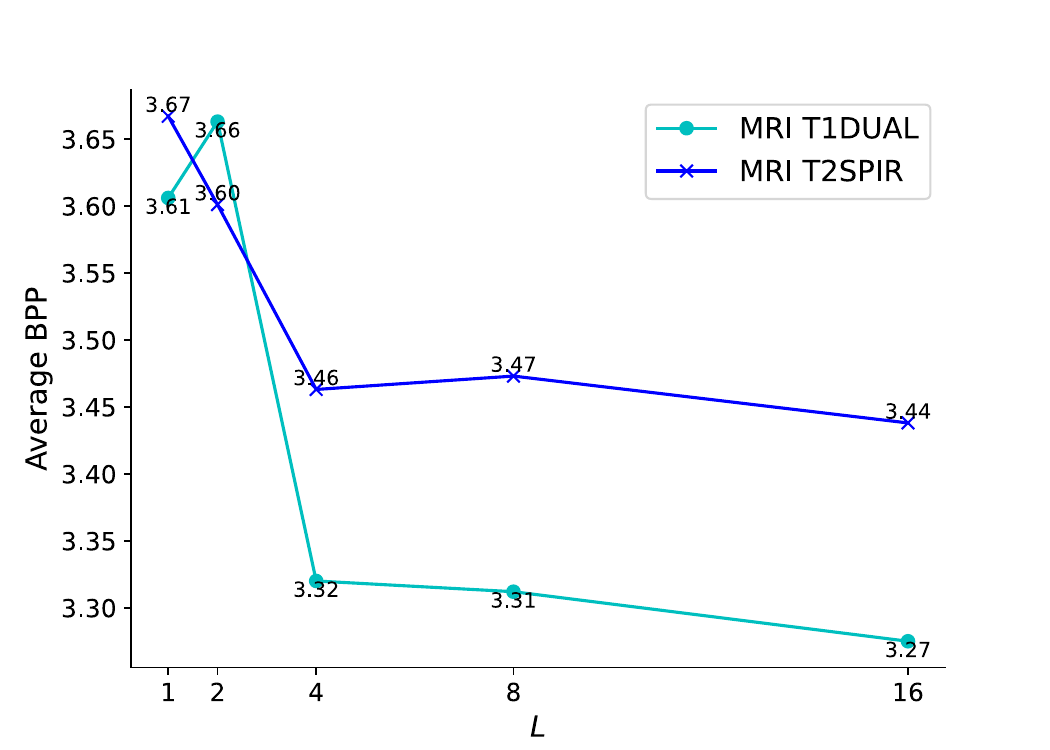}
    \centering
    \caption{Average BPP Results on the CHAOS-MRI datasets with The different scaling factor $L$.}
    \label{fig:L}
\end{figure}

Section \ref{sec:logistic} introduces the variable-scale discrete logistic function, which employs a scaling factor $L$ to mitigate the issue of gradient vanishing.
To assess the impact of this hyperparameter, we conducted several experiments on the $12$-bit CHAOS-MRI dataset.
Specifically, we conducted experiments on the MRI-T1DUAL and MRI-T2SPIR datasets, exploring different values of $L$, including ${1, 2, 4, 8, 16}$.
This means $L$ benefits model training, preventing training from stopping prematurely at the local optimum.
However, as $L$ increases from $4$ to $16$, the decrease in the average BPP curve gradually stabilizes.

\subsection{Hardware Deployment}
We chose the Xilinx Alevo U250 as the evaluation platform for algorithm implementation.
\tabref{tab:res} summarizes the synthesis results of SR-LVC, including floating-point precision, clock frequency, and hardware resource utilization, which consist of the number of Look-Up Tables (LUTs), Flip-Flops (FFs), Digital Signal Processors (DSPs), and Block RAMs (BRAMs).

% =============================================================
\renewcommand{\arraystretch}{1.5}
\begin{table*}
\caption{Implementation Summary of SR-LVC on MPSoC Embedded Platform}
\label{tab:res}
    \begin{center}
    \begin{tabular}{cccccc}
    \hline
    Image Size  & Frequency (MHz) & LUTs            & FFs           & DSPs           & BRAMs (18Kb) \\ \hline
    $256\times256$    & 237             & 123183 (7.13\%) & 74031 (2.14\%) & 1174 (9.55\%) & 503 (9.36\%) \\
    $768\times768$      & 231             & 123825 (7.17\%) & 76120 (2.20\%) & 1174 (9.55\%)  & 503 (9.36\%) \\ \hline
    \end{tabular}
    \end{center}
\end{table*}
%=============================================================

We categorize the maximum input image sizes into two cases: $256\times256$ and $768\times768$, with a floating-point precision set to $16$ bits, following the implementation details outlined in Section \ref{sec:deploy}.
\tabref{tab:res} shows that as the input data volume increases, the number of BRAMs partitioned per array remains constant.
Each 18Kb BRAM stores more valid information and has a higher occupancy rate, resulting in hardware resource utilization remaining constant with the maximum input size.

Additionally, \tabref{tab:delay} presents the resource utilization and performance breakdown for each module within the implementation solution with an input dimension of $256\times256$.
We allocate appropriate DSP computation resources to each module and instantiate the correct number of processing elements (PEs), ensuring that these computation modules exhibit similar processing latencies.
Considering the pipelining mechanism, the total latency for compressing a slice of size $256\times256$ can be represented by the maximum latency among these modules, with Arithmetic Coding running on the CPU taking $26$ms.

% =============================================================
\setlength{\arrayrulewidth}{0.4mm}
\renewcommand{\arraystretch}{1.5}
\begin{table}
\caption{Performance and Resources Utilization of Computation Modules for $256\times256$ Images}
\label{tab:delay}
    \begin{center}
    \begin{tabular}{ccccc}
    \hline
    Module            & \begin{tabular}[c]{@{}c@{}}Processing Latency \\ (ms)\end{tabular} & LUTs  & FFs   & DSPs \\ \hline
    Standard CNN     & 7.73                                                                    & 7673  & 6714  & 297  \\
    Masked CNN       & 7.45                                                                    & 5848  & 5154  & 171  \\
    DSC Module        & 7.83                                                                    & 61646 & 7487  & 662  \\
    Fusion Gate  & 6.73                                                                    & 2154  & 4128  & 10   \\
    Prob. Estimator    & 0.27                                                                    & 471   & 823   & 32   \\ \hline
    Arithmetic Coding & 26                                                                      & -     & -     & -    \\ \hline
    Total             & 26                                                                      & 77792 & 24306 & 1172 \\ \hline
    \end{tabular}
    \end{center}
\end{table}
% =============================================================

\subsection{Compression Speed}
Finally, we compare the compression speed of different compressors on the CPU platform.
\tabref{tab:speed} compares the speeds when compressing a single slice of the MRNet dataset.
The traditional compressor and SR-LVC were tested on the Intel(R) Xeon(R) Gold 6128 CPU, running at $3.4$GHz.
For SR-LVC, $8$ cores and $16$ threads are utilized for parallel computation between different convolution kernels.
L3C and ICEC use the $6$-core Intel i7-6850K CPU, running at $3.8$GHz.
aiWave uses the $4$-core Intel i5-8265U CPU, running at $1.6$GHz.

% ==============================================================
\setlength{\arrayrulewidth}{0.4mm}
\renewcommand{\arraystretch}{1.5}
\begin{table}
\caption{Encoding Time (in Seconds) of Different Methods When Compressing a Single Slice of the MRNet Dataset}.
\label{tab:speed}
    \begin{center}
        \begin{tabular}{cc|cc}
        \hline
        Method & Time (s) & Method  & Time (s) \\ \hline
        PNG          & 0.012  & FFV1      & 0.004                \\
        JP2K         & 0.041                & L3C                               & 0.536                \\
        JP3D         & 0.035  & ICEC         & 1.155                \\
        HEVC         & 0.123   & aiWave         & 986.966              \\ \hline
        SR-LVC (CPU) & 0.086   & SR-LVC (FPGA) & 0.026             \\ \hline
        \end{tabular}
    \end{center}
\end{table}

% =============================================================

In comparison to the learning-based compressor, SR-LVC demonstrates a substantial compression speed advantage, being $13.4$ times faster than ICEC.
Additionally, as SR-LVC can be deployed on the FPGA platform, the compression speed is further enhanced, reaching an increase of nearly $44$ times compared to ICEC, which cannot be deployed on FPGA.
This also demonstrates the applicability of SR-LVC across multiple hardware platforms.
Additionally, due to the pixel-wise latency of SR-LVC, it can initiate compression as pixels arrive, leading to improved real-time performance.

\section{Conclusion}
In this study, we propose a streaming recurrent learning-based framework for the lossless compression of volumetric images.
We employ various convolution structures and gate mechanisms to extract inter-slice dependencies while ensuring that our methods are hardware-friendly.
Extensive experiments on various benchmark datasets demonstrate that our method surpasses traditional lossless volumetric compressors and state-of-the-art learning-based counterparts.
Furthermore, our method exhibits strong generalization capabilities and competitive compression speed, facilitated by hardware implementation.

% =============================================================
\bibliographystyle{unsrtnat}

\end{document}